\documentclass[10pt,a4paper,english,prd,nofootinbib,notitlepage,superscriptaddress,showkeys]{revtex4-1}
\usepackage{color}
\usepackage{babel}
\usepackage{amsmath}
\usepackage{amssymb}
\usepackage{esint}
\usepackage{hyperref}
\usepackage{breakurl}
\usepackage{cleveref}
\usepackage{ulem}
\usepackage{bbm}

\makeatletter

\@ifundefined{textcolor}{}
{%
 \definecolor{BLACK}{gray}{0}
 \definecolor{WHITE}{gray}{1}
 \definecolor{RED}{rgb}{1,0,0}
 \definecolor{GREEN}{rgb}{0,1,0}
 \definecolor{BLUE}{rgb}{0,0,1}
 \definecolor{CYAN}{cmyk}{1,0,0,0}
 \definecolor{MAGENTA}{cmyk}{0,1,0,0}
 \definecolor{YELLOW}{cmyk}{0,0,1,0}
}

\makeatother

\begin{document}

\title{How can we tell whether dark energy is composed by multiple fields?}

\author{Valeri Vardanyan}

\email{v.vardanyan@thphys.uni-heidelberg.de}

\author{Luca Amendola}

\email{l.amendola@thphys.uni-heidelberg.de}

\affiliation{Institut F$\ddot{u}$r Theoretische Physik, Ruprecht-Karls-Universit$\ddot{a}$t
Heidelberg, Philosophenweg 16, 69120 Heidelberg, Germany}
\begin{abstract}
Dark energy is often assumed to be composed by a single scalar field.
The background cosmic expansion is not sufficient to determine whether
this is true or not. We study multi-field scalar-tensor models with
a general dark matter source and write the observable modified gravity
parameters (effective gravitational constant and anisotropic stress)
in the form of a ratio of polynomials in the Fourier wavenumber $k$
of order $2N$, where $N$ is the number of scalar fields. By comparing
these observables to real data it is in principle possible to determine
the number of dark energy scalar fields coupled to gravity. We also
show that there are no realistic non-trivial cases in which the order
of the polynomials is reduced. 
\end{abstract}
\maketitle

\section{Introduction}

Testing the nature of dark energy and its possible modification of
gravity at very large scales is currently one of the most interesting
research activities in cosmology. Modifications of standard gravity
are often modeled by introducing additional mediating fields in the
gravitational Lagrangian. One of the most well studied example is
the so-called Horndeski theory, which adds to the Einstein-Hilbert
Lagrangian a single scalar field that obeys the most general second
order equation of motion \cite{Horndeski:1974}.

It is however in principle possible that dark energy is actually a
composite state of two or more scalar fields interacting through a
common potential and coupled to gravity with different strengths (see
e.g. \cite{BiGalileon_1,BiGalileon_2,Padilla_et_al,continuation_of_Padilla_1,multi_4}).
Since the only observable at background level is the Hubble function
$H(z)$ as a function of redshift, and since to any $H(z)$ one can
always associate a particular single-field potential (see appendix),
it is impossible to tell whether dark energy is driven by multiple
fields instead of just one, as almost universally assumed. In this
paper we show how linear cosmological perturbations produce instead
distinguishable effects that depend on the number of fields.

As shown in several papers (e.g. \cite{Amendola_Kunz_Sapone_2008,deFelice_2011,Silvestri:2013ne,Clifton2011jh,Baker:2012zs}),
a generic modification of gravity introduces at linear perturbation
level two new functions that depend only on background time-dependent
quantities and, in Fourier space, on the wavenumber $k$. One function,
that we denote here with $Y(t,k)$ (sometimes also called $G_{\mathrm{eff}}$),
modifies the standard Poisson equation, while the second one, $\eta(t,k)$,
the anisotropic stress or tilt, provides the relation between the
two gravity potentials $\Psi,\Phi$. In standard gravity, one has
$Y=\eta=1$.

In the so-called quasi-static regime (i.e. for linear scales that
are below the sound horizon) the two functions $Y,\eta$ take a particularly
simple form in the Horndeski models and can be directly constrained
through observations \cite{DeFelice:2011bh,2013PhRvD..87b3501A}.
This also holds true in some cases \cite{Konnig:2014xva} (see also
\cite{Konnig:2014dna} and \cite{Solomon:2014dua}) of bimetric models
\cite{Hassan:2011hr}.

Since the equation of motion of the Horndeski Lagrangian is second
order, the equations contain at most second order space derivatives
and therefore $k^{2}$ terms in Fourier space. It is then no surprise
that the resulting forms of the functions $Y,\eta$ include indeed
polynomials of second order in $k$ 
\begin{equation}
\eta\equiv-\frac{\Phi}{\Psi}=h_{2}\left(\frac{1+k^{2}h_{4}}{1+k^{2}h_{5}}\right)\,,\,Y\equiv-\frac{k^{2}\Psi}{4\pi Ga^{2}\rho\delta}=h_{1}\left(\frac{1+k^{2}h_{5}}{1+k^{2}h_{3}}\right)\,.\label{eq:etay}
\end{equation}
where the $h_{i}$ functions are time dependent functions that depend
on the specific Horndeski model, $a$ is the scale factor, $\rho$
is the average matter density in the Universe, $\delta$ is the matter
density contrast and $\Phi,\Psi$ are the two perturbation potentials,
to be defined below . Also the combination $Y(1+\eta)$ that appears
in the lensing potential equation has the same structure. As shown
in e.g. \cite{Amendola:2013qna}, the fact that the $k$ structure
is fixed regardless of the specific model (within the Horndeski or
the bimetric gravity class), allows one to combine observations of
weak lensing, redshift distortions and galaxy clustering to constrain
or detect modifications of gravity at cosmological scales in a relatively
model-independent way. It is therefore interesting to inquire about
how general the form of Eq.(\ref{eq:etay}) is.

In \cite{Silvestri:2013ne} (see also \cite{Baker:2012zs}) it has
been shown that models that obey $n$-th order equations of motion
generalize the $k$-structure of $Y,\eta$ so that 
\begin{align}
\eta & =h_{2}\frac{P_{n}^{(1)}(k)}{P_{n}^{(2)}(k)}\,,\,Y=h_{1}\frac{P_{n}^{(2)}(k)}{P_{n}^{(3)}(k)}\,.\label{eq:etay-1}
\end{align}
where $P_{n}^{(1)},P_{n}^{(2)},P_{n}^{(3)}$ are even polynomials
of $n-$th order in $k$ with time-dependent coefficients, normalized
for convenience so that $P_{n}^{(1)}(0)=P_{n}^{(2)}(0)=P_{n}^{(3)}(0)=1$.
This case includes Lagrangians which depend on several scalar fields,
because they also obey a system of second-order differential equations
that is dynamically equivalent to a single higher-order equation.
A Lagrangian built out of several scalar fields is in general stable
provided the interaction potential is bounded from below.

In order to compare the models to observations however one needs the
explicit form of the polynomials: this is the main goal of this paper.
Here we consider the simplest non-trivial scalar tensor theory, namely
a multi-field Brans-Dicke model, parametrized by several coupling
constants $\omega_{i}$ and by a general potential $V(\phi_{1},\phi_{2},...)$.
Moreover, we generalize the Brans-Dicke model also in a different
way: we include a generic fluid matter source with equation of state
$w$ and sound speed $c_{s}$, both of which generally time dependent.
This might turn out useful to compare real data to models in which
dark matter includes hot or warm components.

Finally, we investigate whether there are cases in which the higher-order
$k$ terms in the polynomials in Eq.(\ref{eq:etay-1}) cancel out,
thereby effectively reducing the order of the system and, consequently,
the number of observable scalar fields. We find that this is not possible
unless the fields are decoupled from gravity or the Universe is filled
only with radiation. We argue therefore that the $k$-structure of
$Y,\eta$ is a unique signature of the number of dynamically active
dark energy fields coupled to gravity.

Multi-scalar-field scenarios are being considered in gravitational
physics and cosmology since a while. In the context of modified theories
of gravity such models are studied in e.g. \cite{multi_1,multi_3,multi_2,Amendola:2014kwa},
for string theory inspired multifield dark energy models see e.g.
\cite{Marsh_1,Marsh_2} while for the inflationary context see e.g.
\cite{multi_inflat_1,Liddle:1998jc,Copeland:1999cs,multi_inflat_2}.

\section{Standard Brans-Dicke Theory\label{sec:Standard-Brans-Dicke-Theory}}

Let us first derive the cosmological perturbation equations for the
standard Brans-Dicke theory with a potential. The corresponding action
is:

\begin{equation}
S_{1}=\frac{1}{16\pi G}\int d^{4}x\sqrt{-g}(\phi R-\frac{\omega_{BD}}{\phi}g^{\mu\nu}\partial_{\mu}\phi\partial_{\nu}\phi-V(\phi))+\int d^{4}x\sqrt{-g}\mathcal{L}_{matter},
\end{equation}
where as matter fluid we consider a general component characterized
by an equation of state $w$ and a sound speed $c_{s}$, both arbitrarily
time dependent. The equation of motion (e.o.m.) emerging from the
variation with respect to the metric is 
\begin{equation}
\phi G_{\mu\nu}+[\square\phi+\frac{1}{2}\frac{\omega_{BD}}{\phi}(\nabla\phi)^{2}]g_{\mu\nu}-\nabla_{\mu}\nabla_{\nu}\phi-\frac{\omega_{BD}}{\phi}\nabla_{\mu}\phi\nabla_{\nu}\phi+\frac{V(\phi)}{2}g_{\mu\nu}=8\pi GT_{\mu\nu},\label{eq:one_field_grav_eom}
\end{equation}
where 
\begin{equation}
T_{\mu\nu}=(\rho+p)u_{\mu}u_{\nu}+pg_{\mu\nu}
\end{equation}
is the stress-energy tensor of the perfect fluid, with $\rho$ being
the energy density, $p$ being the pressure and $u_{\mu}$ the 4-velocity.
We will need the trace of Eq.(\ref{eq:one_field_grav_eom}) which
reads as 
\begin{equation}
-\phi R+3\square\phi+2\frac{\omega_{BD}}{\phi}(\nabla\phi)^{2}-\frac{\omega_{BD}}{\phi}\nabla^{\alpha}\phi\nabla_{\alpha}\phi+2V(\phi)=8\pi GT.\label{eq:one_field_grav_eom_trace}
\end{equation}
The scalar field e.o.m. is 
\begin{equation}
R+2\frac{\omega_{BD}}{\phi}\square\phi-\frac{\omega_{BD}}{\phi^{2}}\partial^{\alpha}\phi\partial_{\alpha}\phi-V_{,\phi}=0,\label{eq:one_field_scalar_eom}
\end{equation}
where the derivative with respect to the field $\phi$ is denoted
as $V_{,\phi}$. Using Eq.(\ref{eq:one_field_grav_eom_trace}) we
obtain

\begin{equation}
(3+2\omega_{BD})\square\phi+2V(\phi)-V_{,\phi}=8\pi GT.\label{eq:one_field_trace}
\end{equation}

From this equation it follows that in the limit $\omega_{BD}\rightarrow\infty$
the scalar field is constant (see e.g. \cite{weinberg_gravit}) and
therefore in this large-$\omega_{BD}$ limit the gravitational equation
of motion in Eq.(\ref{eq:one_field_grav_eom}) coincides with the
Einstein-Hilbert equation of motion. It is interesting to note that
the Brans-Dicke Lagrangian reproduces also the so-called $f(R)$ models
in the limit in which $\omega_{BD}$ vanishes and $\phi=df(R)/dR$
(see e.g. \cite{amendola_book}).

We start deriving now the cosmological perturbation equations for
the Brans-Dicke theory in a spatially flat Friedmann-Lemaitre-Robertson-Walker
(FLRW) metric. We assume a single fluid matter source with general
equation of state and general sound speed. As it is well know, at
first order in the perturbation parameter one can bring the scalar
line element into the form

\begin{equation}
ds^{2}=a^{2}(\tau)[-(1+2\Psi)d\tau^{2}+(1+2\Phi)dx^{i}dx_{i}],
\end{equation}
where $\tau$ is the conformal time, $a$ is the scale factor and
$\Psi,\Phi$ are the scalar potentials. We also decompose the scalar
field into the background sector and the perturbed sector as \textbf{$\phi(t,\vec{x})=\phi(t)+\varphi(t,\vec{x})$}.
After switching to Fourier space we obtain from the $(0,0)$ element
of Eq.(\ref{eq:one_field_grav_eom}) 
\begin{eqnarray}
-3\frac{\mathcal{H}^{2}}{a^{2}}\varphi+\phi\frac{2}{a^{2}}[3\mathcal{H}(\mathcal{H}\Psi-\Phi^{\prime})-k^{2}\Phi]-\frac{k^{2}}{a^{2}}\varphi-\frac{1}{2a^{2}}\frac{\omega_{BD}}{\phi^{2}}\varphi\phi^{\prime}{}^{2}\nonumber \\
+\frac{1}{a^{2}}\frac{\omega_{BD}}{\phi}\varphi{}^{\prime}\phi^{\prime}+\frac{V_{,\phi}}{2}\varphi=-8\pi G\delta\rho,\label{eq:10}
\end{eqnarray}
where $\mathcal{H}=a'/a$ is the conformal Hubble function, the prime
denotes differentiation with respect to the conformal time $\tau$
and $\delta\rho$ denotes the matter density perturbation.

Now we assume the validity of the so called quasistatic approximation
(see e.g. \cite{Amendola_Toccini-Valentini,DeFelice:2011bh}), which
means we consider only subhorizon scales $k^{2}\gg\mathcal{H}^{2}$
and also that the perturbation fields vary slowly enough with time
so that $\Phi,\Phi',\Psi,\Psi',\varphi,\varphi'$ are all negligible
with respect to $\delta\rho$ unless multiplied by $k^{2}$ or by
the effective scalar field mass $M^{2}\equiv V_{,\phi\phi}$. Note
that in case of large $\omega_{BD}$, the terms like the last term
on the first line of Eq. (\ref{eq:10}) is not very large because
all such terms multiply the derivative of the field, which is always
of order of $1/\omega_{BD}$, as we have discussed above. In this
approximation the previous expression becomes 
\begin{equation}
2\phi\frac{k^{2}}{a^{2}}\Phi+\frac{k^{2}}{a^{2}}\varphi-8\pi G\delta\rho=0.
\end{equation}
Perturbing the scalar field e.o.m. we obtain

\begin{eqnarray}
\frac{1}{a^{2}}[2k^{2}\Psi+4k^{2}\Phi-12(\mathcal{H}^{\prime}+\mathcal{H}^{2})\Psi]+6\frac{1}{a^{2}}\Phi^{\prime\prime}-6\frac{\mathcal{H}}{a^{2}}(\Psi^{\prime}-3\Phi^{\prime})-2\frac{\omega_{BD}}{\text{\ensuremath{\phi}}^{2}}\varphi\nabla^{0}\nabla_{0}\text{\ensuremath{\phi}}+\nonumber \\
2\frac{\omega_{BD}}{\text{\ensuremath{\phi}}}\nabla^{i}\nabla_{i}\varphi+2\frac{\omega_{BD}}{\text{\ensuremath{\phi}}}\nabla^{0}\nabla_{0}\varphi-\frac{2}{a^{2}}\frac{\omega_{BD}}{\text{\ensuremath{\phi}}^{3}}\varphi\phi^{\prime2}+\frac{2}{a^{2}}\frac{\omega_{BD}}{\text{\ensuremath{\phi}}^{2}}\varphi^{\prime}\phi^{\prime}-V_{,\phi\phi}\varphi & =0.
\end{eqnarray}
In the quasistatic limit this becomes:

\begin{equation}
4\frac{k^{2}}{a^{2}}\Phi+2\frac{k^{2}}{a^{2}}\Psi-(2\frac{\omega_{BD}}{\phi}\frac{k^{2}}{a^{2}}+M^{2})\varphi=0.
\end{equation}
Finally we use the traceless part of the gravitational e.o.m. to derive
one more perturbation equation: 
\begin{equation}
\phi G_{0}^{0}-3\nabla^{0}\nabla_{0}\phi-2\nabla^{i}\nabla_{i}\phi+\phi R-\frac{1}{2}\frac{\omega_{BD}}{\phi}\nabla^{i}\phi\nabla_{i}\phi-\frac{3}{2}\frac{\omega_{BD}}{\varphi}\nabla^{0}\phi\nabla_{0}\phi-\frac{3}{2}V(\phi)=8\pi G(T_{0}^{0}-T).\label{eq:one_field_grav_zero_zero_minus_trace}
\end{equation}
Recalling that $\delta T_{0}^{0}-\delta T=-\delta\rho-(3c_{s}^{2}-1)\delta\rho=-3c_{s}^{2}\delta\rho$
, where $c_{s}^{2}=\delta p/\delta\rho$ is the sound speed, we get:
\begin{eqnarray}
-3\varphi\frac{\mathcal{H}^{2}}{a^{2}}+\phi[-6\frac{\mathcal{H}}{a^{2}}\Phi^{\prime}+6\frac{\mathcal{H}^{2}}{a^{2}}\Psi-\frac{2}{a^{2}}k^{2}\Phi]+\frac{3}{a^{2}}\varphi{}^{\prime\prime}+2\frac{k^{2}}{a^{2}}\varphi+6\varphi[\frac{\mathcal{H}^{2}}{a^{2}}+\frac{\mathcal{H}^{\prime}}{a^{2}}]\nonumber \\
-\frac{1}{a^{2}}\phi[-2k^{2}\Psi-4k^{2}\Phi+12(\mathcal{H}^{\prime}+\mathcal{H}^{2})\Psi]+6\frac{1}{a^{2}}\phi\Phi^{\prime\prime}-6\frac{\mathcal{H}}{a^{2}}\phi(\Psi^{\prime}-3\Phi^{\prime})\nonumber \\
+\frac{3}{2}\frac{\omega_{BD}}{\phi^{2}}\varphi\nabla^{0}\phi\nabla_{0}\phi-3\frac{\omega_{BD}}{\phi}\nabla^{0}\varphi\nabla_{0}\phi-\frac{3}{2}V_{,\phi}\varphi=-24\pi Gc_{s}^{2}\delta\rho.
\end{eqnarray}
In the quasi-static limit this reduces to 
\begin{equation}
2\phi\frac{k^{2}}{a^{2}}\Phi+2\frac{k^{2}}{a^{2}}\varphi+2\phi\frac{k^{2}}{a^{2}}\Psi=-24\pi Gc_{s}^{2}\delta\rho.
\end{equation}
Collecting all the above-derived perturbation equations together we
have the following system of equations: 
\begin{align}
2\phi\frac{k^{2}}{a^{2}}\Phi+2\frac{k^{2}}{a^{2}}\varphi+2\phi\frac{k^{2}}{a^{2}}\Psi & =-24\pi Gc_{s}^{2}\delta\rho,\label{eq:one_field_pert_eqn1}\\
4\frac{k^{2}}{a^{2}}\Phi+2\frac{k^{2}}{a^{2}}\Psi-(2\frac{\omega_{BD}}{\phi}\frac{k^{2}}{a^{2}}+M^{2})\varphi & =0,\label{eq:one_field_pert_eqn2}\\
2\phi\frac{k^{2}}{a^{2}}\Phi+\frac{k^{2}}{a^{2}}\delta\varphi & =8\pi G\delta\rho.\label{eq:one_field_pert_eqn3}
\end{align}
In case of $c_{s}^{2}=0$, these expressions agree with \cite{deFelice_2011}.
Solving the system and replacing $\delta\rho$ with \textbf{$\delta\equiv\delta\rho/\rho=8\pi G(\delta\rho)/3\Omega_{m}H^{2}$}
we obtain: 
\begin{equation}
k^{2}\Psi=-\frac{3\Omega_{m}\mathcal{H}^{2}\delta}{2}Y,
\end{equation}
where the modified gravity parameter $Y$ is defined to be

\begin{equation}
Y\equiv\frac{1}{\phi}\frac{(1+3c_{s}^{2})M^{2}\phi+\frac{k^{2}}{a^{2}}(4+2\omega_{BD}+6c_{s}^{2}(1+\omega_{BD}))}{M^{2}\phi+\frac{k^{2}}{a^{2}}(3+2\omega_{BD})}.\label{eq:one_field_Y}
\end{equation}
Let us calculate the first two correction terms on top of
the GR case. Since local gravity measurement constrain the Brans-Dicke
coupling parameter to be very large \cite{BD_constraint}, the
expansion of $Y$ up to second order in powers of $1/\omega_{BD}$
gives 
\begin{equation}
Y\backsimeq\frac{(1+3c_{s}^{2})}{\phi}+\frac{1-3c_{s}^{2}}{2\omega_{BD}\phi}+(3c_{s}^{2}-1)\frac{M^{2}\phi+3\frac{k^{2}}{a^{2}}}{4\frac{k^{2}}{a^{2}}\omega_{BD}^{2}\phi}.
\end{equation}
Note that in order for the coupling constant in the zeroth
order Poisson equation to be equal to Newton's constant at the present
time, we should require $\phi(0)=1$. We define the anisotropic
stress as (see e.g. \cite{2013PhRvD..87b3501A}) 
\begin{equation}
\eta\equiv-\frac{\Phi}{\Psi}=\frac{M^{2}\phi+\frac{k^{2}}{a^{2}}(2+3c_{s}^{2}+2\omega_{BD})}{(1+3c_{s}^{2})M^{2}\phi+\frac{k^{2}}{a^{2}}(4+2\omega_{BD}+6c_{s}^{2}(1+\omega_{BD}))}.\label{eq:one_field_eta}
\end{equation}
If we set $M^{2}=\omega_{BD}=c_{s}^{2}=0$, we get $Y=4/3\phi$, $\eta=1/2$,
which, as expected, are the same values as in a standard $f(R)$ model
in the limit of small mass $M$ provided that $\phi=df(R)/dR$. By
comparing those expressions against observations one can in principle
estimate the coefficients in the polynomials and constrain the values
of the matter sound speed and the field mass. However, we can give
a rough estimate for the numerical values of the coefficients at the
present epoch; for that purpose we use that $\phi(0)=1$ from the
above discussion, assume the potential to be of the form $V(\phi)/16\pi G=M^{2}\phi^{2}/32\pi G$
and consider the case of $c_{s}^{2}\sim0$. From the Friedmann equation
we then have $3H_{0}^{2}\sim V(\phi)$ and therefore $M^{2}\sim6H_{0}^{2}$,
where $H_{0}$ is the present value of the Hubble function. Using
those and the fact that at the present epoch we have $a=1$, $\eta$
and $Y$ can be written as: 
\begin{align}
Y=\frac{1+\frac{2k^{2}}{3H_{0}^{2}}}{1+\frac{k^{2}}{2H_{0}^{2}}},\\
\eta=\frac{1+\frac{k^{2}}{3H_{0}^{2}}}{1+\frac{2k^{2}}{3H_{0}^{2}}}.
\end{align}

The second order expansion of $\eta$ in $1/\omega_{BD}$ is

\begin{equation}
\eta\simeq\frac{1}{1+3c_{s}^{2}}+\frac{(3c_{s}^{2}-1)(3c_{s}^{2}+2)}{2\left(3c_{s}^{2}+1\right)^{2}\omega_{BD}}+(9c_{s}^{4}+3c_{s}^{2}-2)\frac{(1+3c_{s}^{2})M^{2}\phi+(4+6c_{s}^{2})\frac{k^{2}}{a^{2}}}{4\left(3c_{s}^{2}+1\right)^{3}\frac{k^{2}}{a^{2}}\omega_{BD}^{2}}.
\end{equation}
Notice that from Eqs. (\ref{eq:one_field_Y}) and (\ref{eq:one_field_eta})
it follows that for $c_{s}^{2}=1/3,$ i.e. for a relativistic fluid,
both $Y$ and $\eta$ are independent of $\omega_{BD}$. This reflects
the fact that under a conformal rescaling the Brans-Dicke theory can
be recast into a ordinary gravity model with field-matter coupling
proportional to the trace of the energy momentum; when this vanishes,
as for a relativistic fluid, the coupling vanishes as well.

\section{Two-Field Brans-Dicke theory\label{sec:Two-Field-Brans-Dicke-theory}}

We move now to the case in which the Lagrangian includes two fields,
both coupled to gravity. We adopt then the following two-field Brans-Dicke
Action

\begin{eqnarray}
S_{2}=\frac{1}{16\pi G}\int d^{4}x\sqrt{-g}[\phi_{1}R-\frac{\omega_{1}}{\phi_{1}}g^{\mu\nu}\partial_{\mu}\phi_{1}\partial_{\nu}\phi_{1}+\phi_{2}R-\frac{\omega_{2}}{\phi_{2}}g^{\mu\nu}\partial_{\mu}\phi_{2}\partial_{\nu}\phi_{2}-V(\phi_{1},\phi_{2})]\nonumber \\
+\int d^{4}x\sqrt{-g}\mathcal{L}_{matter}.
\end{eqnarray}
The gravitational e.o.m. of this theory is

\begin{eqnarray}
(\phi_{1}+\phi_{2})G_{\mu\nu}+[\square(\phi_{1}+\phi_{2})+\frac{1}{2}\frac{\omega_{1}}{\phi_{1}}(\nabla\phi_{1})^{2}+\frac{1}{2}\frac{\omega_{2}}{\phi_{2}}(\nabla\phi_{2})^{2}]g_{\mu\nu}\nonumber \\
-\nabla_{\mu}\nabla_{\nu}(\phi_{1}+\phi_{2})-\frac{\omega_{1}}{\phi_{1}}\nabla_{\mu}\phi_{1}\nabla_{\nu}\phi_{1}-\frac{\omega_{2}}{\phi_{2}}\nabla_{\mu}\phi_{2}\nabla_{\nu}\phi_{2}+\frac{V(\phi_{1},\phi_{2})}{2}g_{\mu\nu}=8\pi GT_{\mu\nu}.\label{eq:two_field_grav_eom}
\end{eqnarray}
Taking the trace we obtain 
\begin{equation}
-(\phi_{1}+\phi_{2})R+3\square(\phi_{1}+\phi_{2})+\frac{\omega_{1}}{\phi_{1}}(\nabla\phi_{1})^{2}+\frac{\omega_{2}}{\phi_{2}}(\nabla\phi_{2})^{2}+2V(\phi_{1},\phi_{2})=8\pi GT.\label{eq:two_field_grav_eom_trace}
\end{equation}
The $\phi_{1}$ and $\phi_{2}$ e.o.m. are 
\begin{equation}
R+2\frac{\omega_{i}}{\phi_{i}}\square\phi_{i}-\frac{\omega_{i}}{\phi_{i}^{2}}\partial^{\alpha}\phi_{i}\partial_{\alpha}\phi_{i}-V_{,\phi_{i}}=0,\:i=1,2.\label{eq:two_field_scalar_eom}
\end{equation}
Summing Eqs.(\ref{eq:two_field_grav_eom_trace}),(\ref{eq:two_field_scalar_eom})
together we obtain: 
\begin{equation}
(3+2\omega_{1})\square\phi_{1}+(3+2\omega_{2})\square\phi_{2}+2V(\phi_{1},\phi_{2})-V_{,\phi_{1}}-V{}_{,\phi_{2}}=8\pi GT.
\end{equation}
Repeating the arguments after Eq.(\ref{eq:one_field_trace}) we conclude
that this theory reduces to standard general relativity in the large-$\omega_{1},\omega_{2}$
limit.

Proceeding as in the previous sectiom, the perturbed (0,0) component
of Eq.(\ref{eq:two_field_grav_eom}) in the quasi-static approximation
leads to the corresponding perturbation equation: 
\begin{equation}
2(\phi_{1}+\phi_{2})\frac{k^{2}}{a^{2}}\Phi+\frac{k^{2}}{a^{2}}(\varphi_{1}+\varphi_{2})-8\pi G\delta\rho=0.
\end{equation}
Similarly, from the scalar field e.o.m. Eqs.(\ref{eq:two_field_scalar_eom})
we derive the corresponding perturbation equations: 
\begin{equation}
4\frac{k^{2}}{a^{2}}\Phi+2\frac{k^{2}}{a^{2}}\Psi-[2\frac{\omega_{i}}{\phi_{i}}\frac{k^{2}}{a^{2}}+M_{i}^{2}]\varphi_{i}=0,\:i=1,2
\end{equation}
where $M_{i}^{2}\equiv V_{,\phi_{i}\phi_{i}},\:i=1,2$. Perturbing
also the (0,0) component of the traceless part of the gravitational
e.o.m. in the quasistatic limit, we obtain the following system of
perturbation equations: 
\begin{align}
2(\phi_{1}+\phi_{2})\frac{k^{2}}{a^{2}}\Phi+2\frac{k^{2}}{a^{2}}(\varphi_{1}+\varphi_{2})+2(\phi_{1}+\phi_{2})\frac{k^{2}}{a^{2}}\Psi & =-24\pi Gc_{s}^{2}\delta\rho,\label{eq:drho}\\
4\frac{k^{2}}{a^{2}}\Phi+2\frac{k^{2}}{a^{2}}\Psi-[2\frac{\omega_{i}}{\phi_{i}}\frac{k^{2}}{a^{2}}+M_{i}^{2}]\varphi_{i} & =0,\:i=1,2\\
2(\phi_{1}+\phi_{2})\frac{k^{2}}{a^{2}}\Phi+\frac{k^{2}}{a^{2}}(\varphi_{1}+\varphi_{2}) & =8\pi G\delta\rho.
\end{align}

Solving this system and replacing $\delta\rho$ with \textbf{$\delta\equiv\delta\rho/\rho=8\pi G(\delta\rho)/3\Omega_{m}H^{2}$}
as before, we obtain finally

\begin{equation}
k^{2}\Psi=-\frac{3\Omega_{m}\mathcal{H}^{2}\delta}{2}Y,
\end{equation}
where 
\begin{equation}
Y\equiv A_{1}\frac{1+A_{2}k^{2}+A_{3}k^{4}}{1+A_{4}k^{2}+A_{5}k^{4}},
\end{equation}
with the coefficients being

\begin{eqnarray}
A_{1} & = & \frac{1+3c_{s}^{2}}{\phi_{1}+\phi_{2}},\\
A_{2} & = & \frac{2}{\left(3c_{s}^{2}+1\right)(\phi_{1}+\phi_{2})a^{2}}[\frac{(1+3c_{s}^{2})(\phi_{1}+\phi_{2})\omega_{1}+\left(2+3c_{s}^{2}\right)\phi_{1}}{M_{1}^{2}\phi_{1}}\nonumber \\
 & + & \frac{\left(1+3c_{s}^{2}\right)(\phi_{1}+\phi_{2})\omega_{2}+\left(2+3c_{s}^{2}\right)\phi_{2}}{M_{2}^{2}\phi_{2}}],\\
A_{3} & = & \frac{4(2+3c_{s}^{2})(\omega_{2}\phi_{1}+\omega_{1}\phi_{2})+4\omega_{1}\omega_{2}(\phi_{1}+\phi_{2})(1+3c_{s}^{2})}{\left(1+3c_{s}^{2}\right)M_{1}^{2}M_{2}^{2}\phi_{1}\phi_{2}(\phi_{1}+\phi_{2})a^{4}},\label{eq:A3}\\
A_{4} & = & \frac{1}{(\phi_{1}+\phi_{2})a^{2}}\left[\frac{2\omega_{1}(\phi_{1}+\phi_{2})+3\ensuremath{\phi_{1}}}{M_{1}^{2}\phi_{1}}+\frac{2\omega_{2}(\phi_{1}+\phi_{2})+3\ensuremath{\phi_{2}}}{M_{2}^{2}\phi_{2}}\right],\\
A_{5} & = & \frac{4\omega_{1}\omega_{2}(\phi_{1}+\phi_{2})+6(\text{\ensuremath{\phi_{1}}}\omega_{2}+\phi_{2}\omega_{1})}{M_{1}^{2}M_{2}^{2}\phi_{1}\phi_{2}(\phi_{1}+\phi_{2})a^{4}}.\label{eq:A5}
\end{eqnarray}
Considering the limit of $\omega_{1}$ and $\omega_{2}$ both
much larger than unity and keeping the first terms of expansion which
are non-trivial in $k^{2}/a^{2}$, we find for the effective gravitational
constant the result:

\begin{equation}
Y\backsimeq\frac{3c_{s}^{2}+1}{\phi_{1}+\phi_{2}}+\frac{(1-3c_{s}^{2})(\omega_{1}\phi_{2}+\omega_{2}\phi_{1})}{2\omega_{1}\omega_{2}(\phi_{1}+\phi_{2})^{2}}+(3c_{s}^{2}-1)\frac{(\text{\ensuremath{\phi_{1}}}+\phi_{2})\left(M_{1}^{2}\omega_{2}^{2}\text{\ensuremath{\phi_{1}^{2}}}+M_{2}^{2}\omega_{1}^{2}\phi_{2}^{2}\right)+3\frac{k^{2}}{a^{2}}(\omega_{1}\phi_{2}+\omega_{2}\text{\ensuremath{\phi_{1}}})^{2}}{4\frac{k^{2}}{a^{2}}\omega_{1}^{2}\omega_{2}^{2}(\text{\ensuremath{\phi_{1}}}+\phi_{2})^{3}}.
\end{equation}
Similarly, the anisotropic stress is

\begin{equation}
\eta\equiv-\frac{\Phi}{\Psi}=B_{1}\frac{1+B_{2}k^{2}+B_{3}k^{4}}{1+B_{4}k^{2}+B_{5}k^{4}},
\end{equation}
with the coefficients defined as 
\begin{eqnarray}
B_{1} & = & \frac{1}{\left(3c_{s}^{2}+1\right)},\\
B_{2} & = & \frac{1}{(\phi_{1}+\phi_{2})a^{2}}[\frac{\left(\left(3c_{s}^{2}+2\right)\phi_{2}+2\omega_{2}(\phi_{1}+\phi_{2})\right)}{M_{2}^{2}\phi_{2}}+\frac{\left(\text{\ensuremath{\phi_{1}}}\left(3c_{s}^{2}+2\right)+2\omega_{1}(\phi_{1}+\phi_{2})\right)}{M_{1}^{2}\phi_{1}}],\\
B_{3} & = & \frac{2\left(2+3c_{s}^{2}\right)(\omega_{1}\phi_{2}+\omega_{2}\phi_{1})+4\omega_{1}\omega_{2}(\phi_{1}+\phi_{2})}{M_{1}^{2}M_{2}^{2}\phi_{1}\phi_{2}(\phi_{1}+\phi_{2})a^{4}},\label{eq:B3}\\
B_{4} & = & A_{2}=\frac{2}{\left(3c_{s}^{2}+1\right)(\phi_{1}+\phi_{2})a^{2}}[\frac{(1+3c_{s}^{2})(\phi_{1}+\phi_{2})\omega_{1}+\left(2+3c_{s}^{2}\right)\phi_{1}}{M_{1}^{2}\phi_{1}}\nonumber \\
 &  & +\frac{\left(1+3c_{s}^{2}\right)(\phi_{1}+\phi_{2})\omega_{2}+\left(2+3c_{s}^{2}\right)\phi_{2}}{M_{2}^{2}\phi_{2}}],\\
B_{5} & = & A_{3}=\frac{4(2+3c_{s}^{2})(\omega_{2}\phi_{1}+\omega_{1}\phi_{2})+4\omega_{1}\omega_{2}(\phi_{1}+\phi_{2})(1+3c_{s}^{2})}{\left(1+3c_{s}^{2}\right)M_{1}^{2}M_{2}^{2}\phi_{1}\phi_{2}(\phi_{1}+\phi_{2})a^{4}}.
\end{eqnarray}
As expected, we find again that $Y$ and $\eta$ are independent of
$\omega_{1,2}$ if $c_{s}^{2}=1/3$. Notice also that since $A_{2}=B_{4}$
and $A_{3}=B_{5}$, the lensing potential $\Psi-\Phi$ obeys the equation

\begin{equation}
k^{2}(\Psi-\Phi)=-\frac{3}{2}Y(1+\eta)\Omega_{m}\mathcal{H}^{2}\delta,
\end{equation}
in which the combination $Y(1+\eta)$ has in general the same $k$-structure
as $Y$ and $\eta$. However, for the particular case here studied,
i.e. the simple Brans-Dicke model, the null geodesics must remain
the same as in General Relativity and therefore $Y(1+\eta)$ should
be independent of $k$. In fact we find $Y(1+\eta)=(2+3c_{s}^{2})/(\phi_{1}+\phi_{2})$.
This shows that from weak lensing alone it would be impossible to
detect the presence of several Brans-Dicke fields.

In the large $\omega_{1}$ and $\omega_{2}$ limit, again
keeping the first terms which are non-trivial in $k^{2}/a^{2}$, we
obtain for $\eta$

\begin{align}
\eta\backsimeq\frac{1}{3c_{s}^{2}+1}+\frac{(3c_{s}^{2}-1)(3c_{s}^{2}+2)(\omega_{1}\phi_{2}+\omega_{2}\text{\ensuremath{\phi_{1}}})}{2\left(3c_{s}^{2}+1\right)^{2}\omega_{1}\omega_{2}(\text{\ensuremath{\phi_{1}}}+\phi_{2})}-\nonumber \\
\left(9c_{s}^{4}+3c_{s}^{2}-2\right)\frac{\left(3c_{s}^{2}+1\right)(\text{\ensuremath{\phi_{1}}}+\phi_{2})\left(M_{1}^{2}\omega_{2}^{2}\text{\ensuremath{\phi_{1}^{2}}}+M_{2}^{2}\omega_{1}^{2}\phi_{2}^{2}\right)+2\left(3c_{s}^{2}+2\right)\frac{k^{2}}{a^{2}}(\omega_{1}\phi_{2}+\omega_{2}\text{\ensuremath{\phi_{1}}})^{2}}{4(3c_{s}^{2}+1)^{3}\frac{k^{2}}{a^{2}}\omega_{1}^{2}\omega_{2}^{2}(\text{\ensuremath{\phi_{1}}}+\phi_{2})^{2}}.
\end{align}

It is also interesting to compare our two-field results with the standard
single-field ones in the limit when one of the Brans-Dicke parameters
is very large, i.e. when the corresponding field effectively decouples
from gravity. Keeping only the zeroth order term in such an expansion
and setting $\phi_{2}$ to zero we have:

\begin{align}
Y & \backsimeq\frac{\left(1+3c_{s}^{2}\right)M_{1}^{2}\text{\ensuremath{\phi_{1}}}+2\frac{k^{2}}{a^{2}}\left(3c_{s}^{2}(1+\omega_{1})+\omega_{1}+2\right)}{\text{\ensuremath{\phi_{1}}}\left(M_{1}^{2}\text{\ensuremath{\phi_{1}}}+\frac{k^{2}}{a^{2}}(2\omega_{1}+3)\right)},\\
\eta & \backsimeq\frac{M_{1}^{2}\text{\ensuremath{\phi_{1}}}+\frac{k^{2}}{a^{2}}\left(2+3c_{s}^{2}+2\omega_{1}\right)}{\left(1+3c_{s}^{2}\right)M_{1}^{2}\text{\ensuremath{\phi_{1}}}+2\frac{k^{2}}{a^{2}}\left(3c_{s}^{2}(\omega_{1}+1)+\omega_{1}+2\right)}.
\end{align}
As we see they identically coincide with Eqs.(\ref{eq:one_field_Y})
and (\ref{eq:one_field_eta}). In this limit, therefore, the two-field
Lagrangian effectively reduces to a single-field one.

\section{Multi-Field Brans-Dicke theory\label{sec:Multi-Field-Brans-Dicke-theory}}

Finally, we can generalize the previous steps to the $N$-field Brans-Dicke.
By analyzing the previous perturbation equations we can immediately
write down the $N$-field quasi-static perturbation equations as follows:
\begin{eqnarray}
2\sum_{i=1}^{N}\phi_{i}\frac{k^{2}}{a^{2}}\Phi+2\frac{k^{2}}{a^{2}}\sum_{i=1}^{N}\varphi_{i}+2\sum_{i=1}^{N}\phi_{i}\frac{k^{2}}{a^{2}}\Psi & = & -24\pi Gc_{s}^{2}\delta\rho,\label{eq:N_field_1}\\
4\frac{k^{2}}{a^{2}}\Phi+2\frac{k^{2}}{a^{2}}\Psi-[2\frac{\omega_{i}}{\phi_{i}}\frac{k^{2}}{a^{2}}+M_{i}^{2}]\varphi_{i} & = & 0,\:i=1,2,...,N,\label{eq:N_field_2}\\
2\sum_{i=1}^{N}\phi_{i}\frac{k^{2}}{a^{2}}\Phi+\frac{k^{2}}{a^{2}}\sum_{i=1}^{N}\varphi_{i} & = & 8\pi G\delta\rho.\label{eq:N_field_3}
\end{eqnarray}
From Eqs.(\ref{eq:N_field_2}) we have

\begin{equation}
\sum_{i=1}^{N}\varphi_{i}=2\frac{k^{2}}{a^{2}}(2\Phi+\Psi)\sum_{i=1}^{N}\frac{1}{2\frac{\omega_{i}}{\phi_{i}}\frac{k^{2}}{a^{2}}+M_{i}^{2}}.
\end{equation}
Therefore Eqs.(\ref{eq:N_field_1}) and (\ref{eq:N_field_3}) read
respectively

\begin{eqnarray}
2\sum_{i=1}^{N}\phi_{i}\frac{k^{2}}{a^{2}}\Phi+4\frac{k^{4}}{a^{4}}(2\Phi+\Psi)\sum_{i=1}^{N}\frac{1}{2\frac{\omega_{i}}{\phi_{i}}\frac{k^{2}}{a^{2}}+M_{i}^{2}}+2\sum_{i=1}^{N}\phi_{i}\frac{k^{2}}{a^{2}}\Psi & = & -24\pi Gc_{s}^{2}\delta\rho,\\
2\sum_{i=1}^{N}\phi_{i}\frac{k^{2}}{a^{2}}\Phi+2\frac{k^{4}}{a^{4}}(2\Phi+\Psi)\sum_{i=1}^{N}\frac{1}{2\frac{\omega_{i}}{\phi_{i}}\frac{k^{2}}{a^{2}}+M_{i}^{2}} & = & 8\pi G\delta\rho.
\end{eqnarray}
Solving this system one finds 
\begin{eqnarray}
\eta & = & \frac{\sum_{i=1}^{N}\phi_{i}+\left(2+3c_{s}^{2}\right)\frac{k^{2}}{a^{2}}\sum_{i=1}^{N}\frac{1}{2\frac{\omega_{i}}{\phi_{i}}\frac{k^{2}}{a^{2}}+M_{i}^{2}}}{\left(1+3c_{s}^{2}\right)\sum_{i=1}^{N}\phi_{i}+2\left(2+3c_{s}^{2}\right)\frac{k^{2}}{a^{2}}\sum_{i=1}^{N}\frac{1}{2\frac{\omega_{i}}{\phi_{i}}\frac{k^{2}}{a^{2}}+M_{i}^{2}}},\label{eq:eta_N_field}\\
Y & = & \frac{\left(1+3c_{s}^{2}\right)\sum_{i=1}^{N}\phi_{i}+2\left(2+3c_{s}^{2}\right)\frac{k^{2}}{a^{2}}\sum_{i=1}^{N}\frac{1}{2\frac{\omega_{i}}{\phi_{i}}\frac{k^{2}}{a^{2}}+M_{i}^{2}}}{(\sum_{i=1}^{N}\phi_{i})^{2}+3(\sum_{i=1}^{N}\phi_{i})\sum_{i=1}^{N}\frac{1}{2\frac{\omega_{i}}{\phi_{i}}\frac{k^{2}}{a^{2}}+M_{i}^{2}}\frac{k^{2}}{a^{2}}}.\label{eq:Y_N_field}
\end{eqnarray}
The expression for $\eta$ can now be written as

\begin{equation}
\eta=C_{0}\frac{1+\sum_{i=1}^{N}C_{i}k^{2i}}{1+\sum_{i=1}^{N}D_{i}k^{2i}},
\end{equation}
by explicitly specifying the coefficients

\begin{equation}
C_{0}=\frac{1}{1+3c_{s}^{2}},
\end{equation}
and

\begin{eqnarray}
C_{d}=\frac{1}{a^{2d}C_{\star}}[(\sum_{i=1}^{N}\phi_{i})2^{d}\sum_{i_{1}>i_{2}>...>i_{d}}\frac{\omega_{i_{1}}}{\phi_{i_{1}}}\frac{\omega_{i_{2}}}{\phi_{i_{2}}}...\frac{\omega_{i_{d}}}{\phi_{i_{d}}}\prod_{j\neq i_{1},i_{2},...,i_{d}}M_{j}^{2} & +\nonumber \\
\left(2+3c_{s}^{2}\right)2^{d-1}\sum_{i=1}^{N}\sum_{\substack{i_{j}\not=i,\\
j=1,...,d-1\\
i_{1}>i_{2}>...>i_{d-1}
}
}\frac{\omega_{i_{1}}}{\phi_{i_{1}}}\frac{\omega_{i_{2}}}{\phi_{i_{2}}}...\frac{\omega_{i_{d-1}}}{\phi_{i_{d-1}}}\prod_{j\neq i,i_{1},i_{2},...,i_{d-1}}M_{j}^{2}],\label{eq:N_field_coeff}\\
D_{d}=\frac{1}{a^{2d}D_{\star}}[\left(1+3c_{s}^{2}\right)(\sum_{i=1}^{N}\phi_{i})2^{d}\sum_{i_{1}>i_{2}>...>i_{d}}\frac{\omega_{i_{1}}}{\phi_{i_{1}}}\frac{\omega_{i_{2}}}{\phi_{i_{2}}}...\frac{\omega_{i_{d}}}{\phi_{i_{d}}}\prod_{j\neq i_{1},i_{2},...,i_{d}}M_{j}^{2} & + & \mbox{\ensuremath{}}\nonumber \\
2\left(2+3c_{s}^{2}\right)2^{d-1}\sum_{i=1}^{N}\sum_{\substack{i_{j}\not=i,\\
j=1,...,d-1\\
i_{1}>i_{2}>...>i_{d-1}
}
}\frac{\omega_{i_{1}}}{\phi_{i_{1}}}\frac{\omega_{i_{2}}}{\phi_{i_{2}}}...\frac{\omega_{i_{d-1}}}{\phi_{i_{d-1}}}\prod_{j\neq i,i_{1},i_{2},...,i_{d-1}}M_{j}^{2}],
\end{eqnarray}
with $d=1,2,...,N$. Here and in the following expressions by the
second summation sign in the second line of Eq.(\ref{eq:N_field_coeff})\textbf{
}we mean multiple sum over the indices $i_{j},\:j=1,...,d-1$, where
each of $i_{j}$ runs from $1$ to $N$ omitting the value $i$. In
the case of $d=1$ it is meant that there is no summation at all.
The products of $M_{j}^{2}$ are over index $j$ running from $1$
to $N$ omitting the corresponding indices mentioned under the product
signs. $C_{\star}$ and $D_{\star}$ are given by: 
\begin{eqnarray}
C_{\star} & = & (\sum_{i=1}^{N}\phi_{i})\prod_{j=1}^{N}M_{j}^{2},\\
D_{\star} & = & \left(1+3c_{s}^{2}\right)(\sum_{i=1}^{N}\phi_{i})\prod_{j=1}^{N}M_{j}^{2}.
\end{eqnarray}

Similarly, we find

\begin{equation}
Y=\bar{C}_{0}\frac{1+\sum_{i=1}^{N}\bar{C}_{i}k^{2i}}{1+\sum_{i=1}^{N}\bar{D}_{i}k^{2i}},
\end{equation}
where

\begin{equation}
\bar{C}_{0}=\frac{1+3c_{s}^{2}}{\sum_{i=1}^{N}\phi_{i}},
\end{equation}
and

\begin{eqnarray}
\bar{C}_{d}=\frac{1}{a^{2d}\bar{C}_{\star}}[\left(1+3c_{s}^{2}\right)(\sum_{i=1}^{N}\phi_{i})2^{d}\sum_{i_{1}>i_{2}>...>i_{d}}\frac{\omega_{i_{1}}}{\phi_{i_{1}}}\frac{\omega_{i_{2}}}{\phi_{i_{2}}}...\frac{\omega_{i_{d}}}{\phi_{i_{d}}}\prod_{j\neq i_{1},i_{2},...,i_{d}}M_{j}^{2} & +\nonumber \\
2\left(2+3c_{s}^{2}\right)2^{d-1}\sum_{i=1}^{N}\sum_{\substack{i_{j}\not=i,\\
j=1,...,d-1\\
i_{1}>i_{2}>...>i_{d-1}
}
}\frac{\omega_{i_{1}}}{\phi_{i_{1}}}\frac{\omega_{i_{2}}}{\phi_{i_{2}}}...\frac{\omega_{i_{d-1}}}{\phi_{i_{d-1}}}\prod_{j\neq i,i_{1},i_{2},...,i_{d-1}}M_{j}^{2}],\\
\bar{D}_{d}=\frac{1}{a^{2d}\bar{D}_{\star}}[(\sum_{i=1}^{N}\phi_{i})^{2}2^{d}\sum_{i_{1}>i_{2}>...>i_{d}}\frac{\omega_{i_{1}}}{\phi_{i_{1}}}\frac{\omega_{i_{2}}}{\phi_{i_{2}}}...\frac{\omega_{i_{d}}}{\phi_{i_{d}}}\prod_{j\neq i_{1},i_{2},...,i_{d}}M_{j}^{2} & +\nonumber \\
3(\sum_{i=1}^{N}\phi_{i})2^{d-1}\sum_{i=1}^{N}\sum_{\substack{i_{j}\not=i,\\
j=1,...,d-1\\
i_{1}>i_{2}>...>i_{d-1}
}
}\frac{\omega_{i_{1}}}{\phi_{i_{1}}}\frac{\omega_{i_{2}}}{\phi_{i_{2}}}...\frac{\omega_{i_{d-1}}}{\phi_{i_{d-1}}}\prod_{j\neq i,i_{1},i_{2},...,i_{d-1}}M_{j}^{2}],
\end{eqnarray}
with $d=1,2,...,N$. $\bar{C}_{\star}$ and $\bar{D}_{\star}$ are
given by: 
\begin{eqnarray}
\bar{C}_{\star} & = & \left(1+3c_{s}^{2}\right)(\sum_{i=1}^{N}\phi_{i})\prod_{j=1}^{N}M_{j}^{2},\\
\bar{D}_{\star} & = & (\sum_{i=1}^{N}\phi_{i})^{2}\prod_{j=1}^{N}M_{j}^{2}.
\end{eqnarray}
Note that $D_{d}=\bar{C}_{d}$ identically for $\forall d$ (see also
\cite{Silvestri:2013ne}).

We do not have any constraint on the values of the coupling
constants, the field masses or the field trajectories, therefore the
above coefficients will depend in general on the specific model and
should be determined by comparison with the experimental data, which
is beyond the scope of this paper.

\section{reducing the polynomials order?}

In this section we discuss under which circumstances the polynomials
in $\eta$ and $Y$ effectively reduce to lower order ones at all
times. If this were possible then the observations (at least at first
order in the perturbations) would not be able to establish the number
of active coupled dark energy fields, not even in principle.

From Eqs.(\ref{eq:eta_N_field}) and (\ref{eq:Y_N_field}) one can
see that the order of the polynomials can be effectively reduced only
in one of the following cases: \emph{$a$}) $c_{s}^{2}=1/3$, \emph{$b$})
at least one of the coupling constants is infinitely large, \emph{$c$})
one or more $M_{i}^{2}\rightarrow\infty$, \emph{$d$}) one or more
$\phi_{i}\rightarrow0$, \emph{$e$}) there are $i$ and $j$ such
that $\omega_{i,j}\rightarrow0$, \emph{f})) there are $i$ and $j$
such that $\frac{\omega_{i}}{\phi_{i}}=\alpha\frac{\omega_{j}}{\phi_{j}}$
and $M_{i}^{2}=\alpha M_{j}^{2}$, where $\alpha$ is a non-zero real
number. Note that $\alpha=-1$ is conceptually different from all
the other values of $\alpha$; in the former case two summands for
each pair $i,j$ disappear from the sum, whereas in the latter case
the number of summands reduces only by one for each pair $i,j$. We
will comment more on this difference below.

As we have already discussed before, in case of $c_{s}^{2}=1/3$,
i.e. radiation, our theory is conformally equivalent to the standard
general relativity and the fields (regardless of their number) do
not have an observable impact except for a time-dependence of the
effective gravitational constant.

The case \emph{$b$}) is the limit when the field with infinite (or
very large) $\omega_{i}$ is completely decoupled from the Brans-Dicke
theory (as an example of this reduction see the end of Section \ref{sec:Two-Field-Brans-Dicke-theory}).
The case \emph{$c$}) implies that the field with infinite (very large)
mass is static and just renormalizes the gravitational constant. Case
\emph{$d$}) is unphysical because it leads to a singular Lagrangian.
In the case \emph{$e$}) the kinetic terms of the $i$ and $j$ fields
are absent from the Lagrangian and the $\phi_{i}R+\phi_{j}R$ term
can be interpreted as a non-minimal coupling of a single field: $\psi_{ij}R=(\phi_{i}+\phi_{j})R$.

Case \emph{$f$}) is non trivial and we need to check the consistency
of this case with the field equations of motion. To do so we substitute
the condition \emph{$f$}) in the $\phi_{i}$ equation of motion (\ref{eq:two_field_scalar_eom}):

\begin{equation}
R^{(0)}+2\alpha\frac{\omega_{j}}{\phi_{j}}\nabla^{0}\nabla_{0}\phi_{i}+2\alpha\frac{\omega_{j}}{\phi_{i}\phi_{j}a^{2}}\phi_{i}^{\prime2}-V_{,\phi_{i}}|_{\phi_{i}=\omega_{i}\phi_{j}/\alpha\omega_{j}}=0,
\end{equation}
where $R^{(0)}$ is the background Ricci scalar. In order for the
above equation to be consistent with the $\phi_{j}$ e.o.m: 
\begin{equation}
R^{(0)}+2\frac{\omega_{j}}{\phi_{j}}\nabla^{0}\nabla_{0}\phi_{j}+2\frac{\omega_{j}}{\phi_{j}^{2}a^{2}}\phi_{j}^{\prime2}-V_{,\phi_{j}}|_{\phi_{i}=\omega_{i}\phi_{j}/\alpha\omega_{j}}=0,
\end{equation}
we need that $\phi_{j}=\alpha\phi_{i}$ (leading to $\omega_{i}=\omega_{j}$)
and $V_{,\phi_{i}}|_{\phi_{i}=\phi_{j}/\alpha}=V_{,\phi_{j}}|_{\phi_{i}=\phi_{j}/\alpha}$.
This means that depending on whether $\alpha=-1$ or $\alpha\neq-1$,
either these two fields decouple completely from the Lagrangian or
we can collect the terms corresponding to $\phi_{i}$ and $\phi_{j}$
in a way that they effectively act as one field.

As an example we can consider a potential without $\phi_{i},\phi_{j}$
interaction: 
\begin{equation}
V(\phi_{1},\phi_{2},...,\phi_{N})=\sum_{k=1}^{D}(\lambda_{k}^{(i)}(\phi_{i})^{k}+\lambda_{k}^{(j)}(\phi_{j})^{k})+V(\phi_{1},...,\hat{\phi_{i}},...,\hat{\phi_{j}},...,\phi_{N}),
\end{equation}
where the coefficients $\lambda_{k}$ are some constant numbers, $D$
is a natural number and hat represents exclusion. For the derivatives
of this potential we have: 
\begin{eqnarray}
V_{,\phi_{i}}|_{\phi_{i}=\phi_{j}/\alpha} & = & \sum_{k=1}^{D}k\lambda_{k}^{(i)}\frac{(\phi_{j})^{k-1}}{\alpha^{k-1}},\\
V_{,\phi_{j}}|_{\phi_{i}=\phi_{j}/\alpha} & = & \sum_{k=1}^{D}k\lambda_{k}^{(j)}(\phi_{j})^{k-1}.
\end{eqnarray}
In order for the two derivatives to be equal one needs $\lambda_{k}^{(i)}=\alpha^{k-1}\lambda_{k}^{(j)}$.
It is easy to see that this condition is consistent with the proportionality
of the field masses: $V_{,\phi_{i}\phi_{i}}|_{\phi_{i}=\phi_{j}/\alpha}=\alpha V_{,\phi_{j}\phi_{j}}|_{\phi_{i}=\phi_{j}/\alpha}$.

To exemplify the $\alpha=-1$ case we study the conditions for $A_{3},A_{5}$
and $B_{3}$ given in Eqs.(\ref{eq:A3}, \ref{eq:A5}, \ref{eq:B3})
to be zero. It is easy to see that they are consistently zero in one
of the following cases: $c_{s}^{2}=\frac{1}{3}$, $\omega_{1}=\omega_{2}=0$,
at least one of $M_{1}^{2}$ and $M_{2}^{2}$ is infinitely large,
at least one of $\phi_{1}$ and $\phi_{2}$ is zero, or, finally,
$\phi_{1}=-\phi_{2}$ and $\omega_{1}=\omega_{2}$. In the latter
case, which can be associated with the $\alpha=-1$ situation above,
both fields completely decouple from the Lagrangian. This sheds more
light on the case \emph{$f$}) above; the cancellation of higher order
terms in this case is possible if a pair of fields decouples completely
from the Lagrangian ($\alpha=-1$), or if they are still active but
in Lagrangian one can group together the terms corresponding to the
considered pair of fields in such a way that they lead to terms resembling
a singe field ($\alpha\neq-1$). It is important to note that even
though the cancellation of higher order terms was shown only at linear
level, the fact that it leads to a lower number of active fields in
the Lagrangian means that this in fact is a non-perturbative effect.

This means that apart from the case \emph{$a$}) all the other cases
lead to an effectively smaller number of active fields and therefore
are trivial, in the sense that there exists a field redefinition that
brings the Lagrangian into a function of a smaller number of fields.
This shows that if dark energy is composed by $N$ fields that are
dynamically active and coupled to gravity, their presence can in principle
be detected in the observables $Y$ and $\eta$. Needless to say,
the effective detectability of multiple dark energy fields depends
on many other factors and is beyond the scope of this paper.

\section{Conclusions\label{sec:Conclusions}}

We investigated multi-field scalar-tensor models of the Brans-Dicke
type, in presence of a generic source characterized by an equation
of state and a sound speed. Our aim was to find the modified gravity
parameters that affect the Poisson equation and the anisotropic stress
in the quasi-static limit. The interest in this approach lies in the
fact that future large scale observations will probe the $k$-structure
of modified gravity \cite{Amendola:2013qna} and possibly detect whether
dark energy is composed by one or several distinct components.

We have shown that multi-field Brans-Dicke models with $N$ scalar
fields are described by the anisotropic stress $\eta$ and the effective
gravitational constant $Y$ as 
\begin{align}
\eta & =h_{2}\frac{P_{n}^{(1)}(k)}{P_{n}^{(2)}(k)}\,,\,Y=h_{1}\frac{P_{n}^{(2)}(k)}{P_{n}^{(3)}(k)}\,.\label{eq:etay-1-1}
\end{align}
where $P_{n}^{(1)},P_{n}^{(2)},P_{n}^{(3)}$ are even polynomials
of order $n=2N$ in $k$, with coefficients that depend on the evolution
of background quantities and on the matter sound speed $c_{s}$. We
also find that there are no realistic non-trivial cases in which the
higher-order $k$ terms cancel out. Each term in the three polynomials
introduces a new characteristic time-dependent scale $k_{n}$, with
$n=2,4,...,2N$, for total of 3$N$ scales, which can in principle
be individually measured and employed to constrain the number of fields
$N$.

An interesting question that naturally arises is how well can future
surveys like e.g. the Euclid satellite \cite{Euclid-r,Euclid_TWG},
probe the existence of more than one scalar field and constrain the
values of the Brans-Dicke parameters $\omega_{i}$. This is left to
future work. 
\begin{acknowledgments}
We acknowledge support from DFG through the project TRR33 ``The Dark
Universe''. We thank Pedro Ferreira and Alessandra Silvestri for
useful discussions. V.V. acknowledges financial support from DAAD.
This paper benefited of discussions within the Euclid Theory Working
Group. 
\end{acknowledgments}

\section*{Appendix}

Here we show that there always exist a single field potential $V(\phi)$
that reproduces any observed Hubble parameter as a function of redshift,
$H(z)$. We assume the Universe contains matter with a known generic
equation of state $w_{m}$ and a single canonical minimally coupled
scalar field. The two independent Einstein equations are (we set $8\pi G=1$)
\begin{align}
3H^{2} & =\rho_{m}+\frac{\dot{\phi}^{2}}{2}+V(\phi)\label{eq:v}\\
2\dot{H} & =-\dot{\phi}^{2}-\rho_{m}(1+w_{m})\label{eq:hd}
\end{align}
Adopting the redshift $z$ as time coordinate, we can write from (\ref{eq:hd})
\begin{equation}
H(z)^{2}(1+z)^{2}\left(\frac{d\phi}{dz}\right)^{2}=2(1+z)H(z)\frac{dH}{dz}-\rho_{m}(z)(1+w_{m}(z))\label{eq:pdd}
\end{equation}
where $\rho_{m}(z)$ is the solution of 
\begin{equation}
(1+z)\frac{d\rho_{m}}{dz}=3\rho_{m}(z)(1+w_{m}(z))
\end{equation}
From Eq.(\ref{eq:pdd}) one can obtain $\phi(z)$. Then by combining
(\ref{eq:pdd}) and (\ref{eq:v}) one obtains 
\begin{equation}
V=3H(z)^{2}-(1+z)H(z)\frac{dH(z)}{dz}+\frac{\rho_{m}(z)}{2}(w_{m}(z)-1)
\end{equation}
which gives $V(z)$. Finally, by inverting $\phi(z)$, it is possible
to reconstruct $V(\phi)$ for any observed $H(z)$. There is no guarantee
however that the formal solution so obtained is stable or unique or
free of singularities.

\bibliography{multiscalar}

\end{document}